\documentclass[twocolumn,secnumarabic,amssymb, nobibnotes, aps, prd, showpacs,amsmath]{revtex4}
\usepackage[dvipdf]{graphicx}

\usepackage{graphicx}
\usepackage{dcolumn}
\usepackage{bm}
\begin{document}
\title{Neuronal networks with coupling through amyloid beta: towards a theory for Alzheimer's disease.}
\author{V. Resmi}
\email{v.resmi@iiserpune.ac.in}
\affiliation{Indian Institute of Science Education and Research, 
Pune - 411021, India} 
\author{G. Ambika}
\email{g.ambika@iiserpune.ac.in}
\affiliation{Indian Institute of Science Education and Research, 
Pune - 411021, India} 
\author{R. E. Amritkar}
\email{amritkar@prl.res.in}
\affiliation{Physical Research Laboratory, Ahmedabad - 380009, India}
\author{G. Rangarajan}
\email{rangaraj@math.iisc.ernet.in}
\affiliation{Indian Institute of Science, Bangalore - 560 012, India}
\begin{abstract}
Alzheimer's disease (AD) is a common form of dementia observed in the elderly due to neurodegenerative disorder and dysfunction. This arises from alterations in synaptic functioning of neurons leading to cognitive impairment and memory loss.  Recent experimental studies indicate that the amyloid beta (A$\beta$) protein, in its dimer and oligomer forms, affects the synaptic activity of neurons in the early stages of AD. However, the precise mechanism underlying A$\beta$ induced synaptic depression is still not clearly understood. In this paper, we introduce an electrical model that provides a possible mechanism for this.  Our studies show that the competing effects of synaptic activity and the indirect interaction mediated by A$\beta$ , can disrupt the synchrony among neurons and severely affect the neuronal activity. This then leads to sub-threshold activity or synaptic silencing. This is in agreement with the reported disruption of cortical synaptic integration in the presence of A$\beta$ in transgenic mice.   We suggest that direct electrophysiological measurements of A$\beta$ activity can establish its role in AD and the possible revival, proposed in our model. The mechanism proposed here is quite general and could account for the role of the relevant protein in other neuronal disorders also. 
\end{abstract}
\pacs{ 05.45.Xt, 87.19.xr, 87.19.lj }
\maketitle
\section{Introduction}
Alzheimer's disease (AD) is the most common cause for dementia characterized by a progressive neuronal dysfunction leading to neuronal death and depressed brain function \cite{Sha09, Sel02, Tan05}. It has been established over the past decade that pathologically this neurodegenerative disorder may arise from extra cellular deposits of Amyloid  $\beta$ (A$\beta$) protein  in senile plaques and inter neuronal neuro-fibrillatory tangles \cite{Sel00}.  However, recent experimental studies indicate that the amyloid beta (A$\beta$) protein, in its dimer and oligomer forms, affects the synaptic activity of neurons in the early stages of AD \cite{Wal07, Wal00, Cle05, Wal02}. The actual mechanism by which A$\beta$ oligomers participates in the impairment of synaptic function is still not clearly established.

In this work, we introduce a physical model that can explain the mechanism of A$\beta$ induced synaptic depression especially in the early stages and point towards possible methods for prevention or reversal. We start with the simple hypothesis that since the primary functions of neurons are through electrical activity, it is quite possible that the disturbance or impairment of this function is also electrical in nature. This is supported by the fact there is no evidence of physical blocking or damage of synapses or neurons in the very early stages \cite{Cle05}. Amyloid beta, being a protein is not expected to have any electrical activity of its own and no such activity has been reported so far. But there exists a possibility of an induced electrical activity (for example, through polarization) caused by electrical signals generated by nearby neurons. Recent molecular dynamics based study on A$\beta$ indicates the possibility of such an induced dipole moment due to rearrangement of charges in the presence of an electric field \cite{Tos09}. This induced electrical activity of A$\beta$, in turn, provides an input back to the surrounding neurons (giving rise to a feedback loop). Thus A$\beta$ can act as an intermediate agency that provides an extra channel of interaction among neurons.  This is in addition to the regular synaptic interaction between the neurons. The normal coupling through synapse leads to in-phase synchronization of the activities of neurons. The additional coupling through A$\beta$  is equivalent to an environmental coupling of the type discussed in a previous paper \cite{Res10}, where it was shown that it can lead to an anti-phase synchronization. Thus the above two types of couplings between the neurons represent two competing synchronizing tendencies.  Both these tendencies become effective above their respective thresholds for the coupling strengths.  In addition, if their effects balance each other, we expect the electrical activity of the neurons to be severely affected.  The numerical simulation carried out on a neuronal network with additional coupling through A$\beta$ establishes disruption of activity and silencing.
\section{Neurons coupled indirectly via A$\beta$}
The mechanism proposed in the paper is applied to a network of 10 neurons using the Hindmarsh-Rose ( HR)  model \cite{Hin84}.  A simple way of modeling the effect of the extra channel via A$\beta$,  in such a network  is given below.
\begin{eqnarray}
\dot{x}_i = y_i - x_{i}^{3} + a x_i^2 - z_i + I  \nonumber \\ 
+ \epsilon_a w + \frac{\epsilon_s (V_r-x_i)}{n_i} \sum_j A_{ij} \frac{1}{1+\exp(-\lambda(x_j-\theta))}  \nonumber \\
\dot{y}_i = 1-b x_i^2 - y_i \nonumber \\
\dot{z}_i = \rho( s (x_i + \chi) - z_i) \nonumber \\
\dot{w} = -\kappa w - \frac{\epsilon_a}{n} \sum_j x_j
\label{eq1}
\end{eqnarray}
The first three equations in (\ref{eq1}) come from the HR model (except for the $w$ term) and represent the neuronal activity in a normal brain. Here the variable $x$ represents the voltage of the neuron and $y$ and $z$ are variables related to ion currents and the neurons are coupled via excitatory synaptic couplings \cite{Bel05} represented by the last term in the first equation.  We model the pathological activity of A$\beta$ in the case of AD by the additional variable $w$. The nature of the electrical activity of A$\beta$ considered here is not inherent but induced. So without input from neurons, this induced electrical activity should decay. This is represented by the decay term, - $\kappa w$, in the $w$ equation where $\kappa$ is the damping parameter. The second term in the $w$ equation represents the input from the surrounding neurons. The additional $\epsilon_a w$ term in $x$ equation is the extra input to the neuron voltage through A$\beta$ as induced electrical activity.  Here,  we choose the parameters $a=3$,$b=5$, $\rho=0.006$, $s=4$, $\chi=1.6$,$I=3.2$, such that the individual neurons have chaotic responses. The damping parameter of the amyloid beta is taken to be  $\kappa=1$ . The synaptic coupling is modeled by the sigmoidal function given in the last term in the first equation. Here the reversal potential $V_r < x_i$ makes the synapse excitatory. The other parameters in the synaptic coupling term are  chosen to be $\theta = -0.25$ and  $\lambda=10$. The network is constructed using an adjacency matrix A, whose elements are randomly chosen as either 0 or 1 in Eq.(1). $A_{ij}=1$  indicates that $i^{th}$ neuron receives synaptic input from $j^{th}$ neuron. Here the matrix A is asymmetric, such that the coupling is mostly unidirectional though bidirectional couplings can also occur. The synaptic input represented by a sigmoid function is the self-feedback current of  $i^{th}$  neuron modulated by the $j^{th}$ neuron. Using random initial condtions, the network is evolved using Runge-Kutta algorithm   with step size $0.01$ for sufficiently long time.

\begin{figure}
\centerline{\includegraphics[width=.50\textwidth]{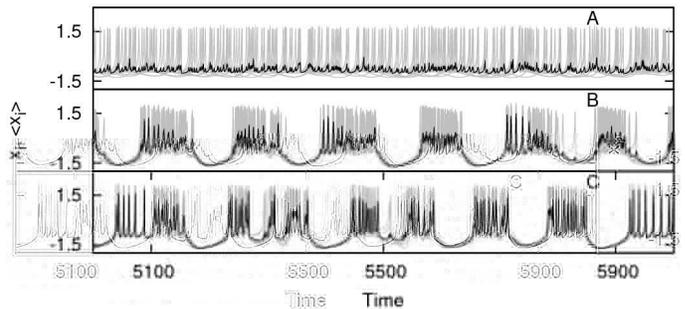}}
\caption{\label{fig1}Time series of the voltage of all neurons (grey) and their average (black). Here a random network of 10 neurons is used with different strengths of synaptic coupling in the absence of coupling through A$\beta$.  (A) Unsynchronized state for $\epsilon_s = 0.01$ . Here the average voltage is much smaller than individual spikes.  (B) Synchronized bursting for  $\epsilon_s = 0.5$ and (C),  Synchronized states for  $\epsilon_s = 1.0$. HR parameters chosen are as $a=3$,$b=5$,$\rho=0.006$,$s=4$,$\chi=1.6$ and $I = 3.2$,  for which individual neurons are in the chaotic state. }
\end{figure}

First we study the baseline activity by considering a network of 10 neurons with HR model as the nodal dynamics and coupled via excitatory synapses alone ( ignoring the influence of A$\beta$). For very small strengths of synaptic coupling $\epsilon_s$, the neurons do not act in synchrony. This is clear from the sequences of neuronal voltages plotted in Fig.~\ref{fig1}.A. Here the grey lines give the neuronal voltages from all the neurons while the black shows the average response. The emergence of synchronous behavior becomes evident as the synaptic strength $\epsilon_s$ increases.  Thus, Figs.~\ref{fig1}.B and \ref{fig1}.C correspond to two typical cases of coordinated activity when only synaptic interaction is present. 
 
\begin{figure*}[ht]
\centerline{\includegraphics[width=.85\textwidth]{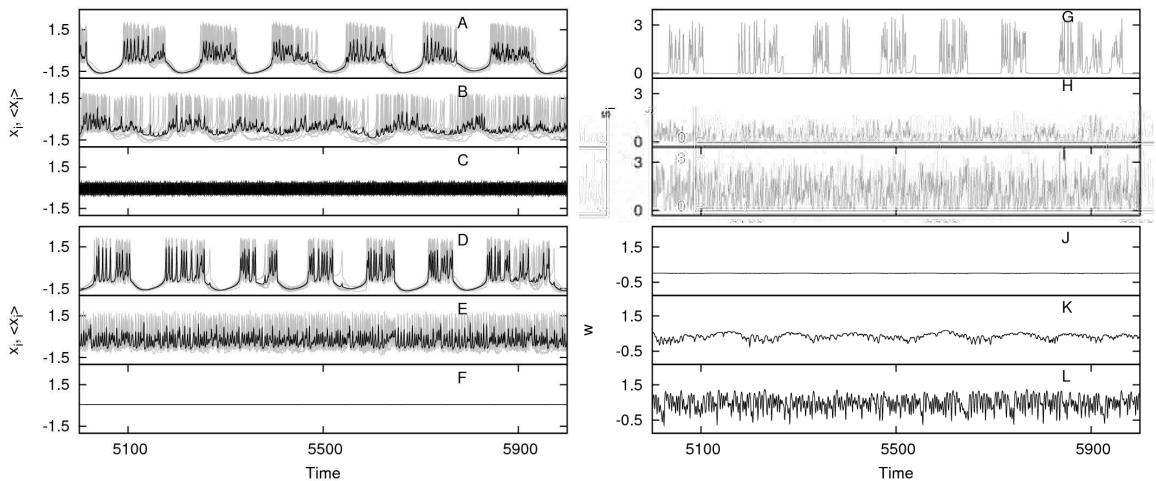}}
\caption{\label{fig2} {\bf Disruption of neuronal activity by A$\beta$}. (A)-(F) Time series of the voltage of all neurons in the network (grey) and their average (black) for different  strengths of synaptic coupling and coupling through A$\beta$. (A) Normal synchronized state ($\epsilon_s = 0.50, \epsilon_a = 0.01$) (B) Desynchronized state ($\epsilon_s = 0.50, \epsilon_a = 0.50$). (C) Sub-threshold activity  ($\epsilon_s = 0.50, \epsilon_a = 3.50$) (D) Normal synchronized state ($\epsilon_s = 1.00, \epsilon_a = 0.01$). (E)  Desynchronized state ($\epsilon_s = 1.0, \epsilon_a = 1.60$) (F) synaptic silencing ($\epsilon_s = 1.00, \epsilon_a = 2.00$). (G-I)  Synaptic input to one neuron and (J-L) Induced activity on A$\beta$ as a function of time at different strengths of synaptic coupling and coupling through A$\beta$  (G) Synaptic input for synchronized state ($ \epsilon_s = 1.00, \epsilon_a = 0.01$) (H) synaptic input for desynchronized activity near sub-threshold oscillatory state ( $\epsilon_s = 0.50, \epsilon_a = 0.50$ ). (I) synaptic input for desynchronized activity near neuronal silencing ($\epsilon_s = 1.0, \epsilon_a = 1.60$) (J) Induced activity on A$\beta$ due to signal from neurons for $\epsilon_s = 1.00, \epsilon_a = 0.01$ (K) Induced activity on A$\beta$  for $\epsilon_s = 0.50, \epsilon_a = 0.50$. (L) Induced activity on A$\beta$ for $\epsilon_s = 1.0, \epsilon_a = 1.60$. }
\end{figure*}

Next, we introduce the extra coupling through A$\beta$ in the same network and the consequent changes in neuronal voltages are shown in Fig. 2.A-F.  We see that the competing effects of the two modes of interaction lead to severe disruptions of the neuronal activity which result in sub threshold behavior and in some cases to complete silencing. Further, the synaptic activity at one neuron is calculated using the last term in Eq.~(\ref{eq1}) and is shown in Fig.~\ref{fig2}.G,H,I. In Fig.~\ref{fig2},  we  plot the total synaptic activity at one neuron  and for different values of A$\beta$ coupling ,show that A$\beta$ coupling also disrupts the synaptic behavior (Fig.2.G-I). We note that such disruptions in synaptic behavior due to A$\beta$ have been reported in earlier experimental observations \cite{Sha08}. The induced activity of A$\beta$ based on our mechanism is evident from the behavior of the variable $w$ shown in Fig 2.J-L.  This induced activity of A$\beta$ is caused by the electrical signal from the neurons and when it is of sufficient strength, it can cause neuronal dysfunction. We also observe that the route to synaptic silencing or sub threshold activity is through a stage of desynchronization of the neuronal outputs. This is clear from Fig.2.B and E where the strength of $\epsilon_a$  is just not sufficient for  synaptic depression. We note that this is consistent with the measurements of cortical neuronal signals reported in transgenic mice \cite{Ste04}.

\begin{figure}
\centerline{\includegraphics[width=0.50\textwidth]{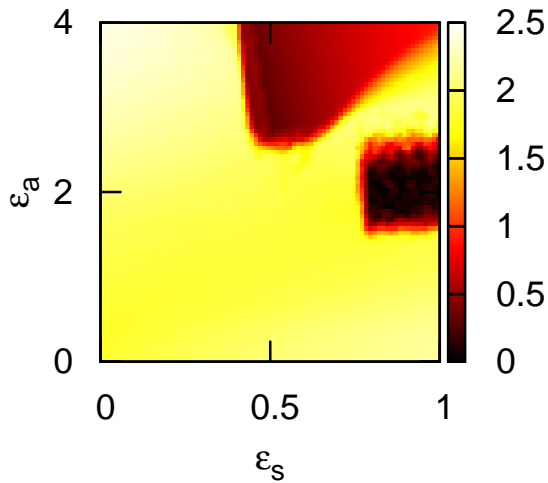}}
\caption{\label{fig3}Regions of disrupted neuronal activity(shown in red and black) in the parameter plane of coupling strengths ($\epsilon_s, \epsilon_a$).The maximum voltage from all neurons , averaged over 10 random realizations of the network of 10 neurons is used as the index and specific nature of activity is verified from time series of a few representative cases. Black regions correspond to regions with total absence of activity and the red regions correspond to regions of sub-threshold activity. Yellow regions correspond to normal neuronal activity which can include unsynchronized bursts and spikes, synchronized bursts, and synchronized spikes for different values of  $\epsilon_a$, and $\epsilon_s$ }
\end{figure}

That the resultant behavior of neurons under the two competing interactions depends on their relative strengths is evident from Fig. 2. Hence we study in detail the neuronal activity in the parameter plane of the coupling strengths ($\epsilon_a$, $\epsilon_s$ ). The maximum value of the voltage of neurons in the network is used as an indicator to identify regions of synaptic silencing in the parameter plane of coupling strengths.  This is shown in Fig.~\ref{fig3}, where black regions correspond to complete electrical inactivity (quiescent state) and red regions correspond to sub-threshold oscillations. We find two patches of red and black regions in the parameter plane where the two competing effects balance each other. 

Based on our model, we propose that one of the possible mechanisms to prevent this regulatory activity of A$\beta$ is either inhibiting electrical activity on A$\beta$ or by reducing the value of parameter $\kappa$ which amounts to very slow relaxation of the induced activity of A$\beta$.  The latter case is shown in Fig. 4, where by decreasing $\kappa$, the synaptic activity is found to be restored.
\begin{figure}
\centerline{\includegraphics[width=0.4\textwidth]{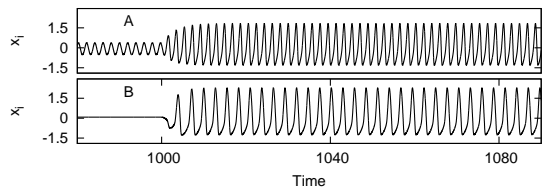}}
\caption{\label{fig4}Revival of neuronal activity by preventing leakage of induced activity of A$\beta$. A,Voltage of a neuron recovering from sub-threshold state with $\epsilon_s = 0.5, \epsilon_a = 3.5$ .B, Voltage of a neuron recovering from silent state with  $\epsilon_s = 1.0$ and $\epsilon_a = 2.0$. In both cases, $\kappa$ is reduced to 1\% of its initial value  at  the $1000^{th}$ time unit.}
\end{figure} 
\section{Discussion}
We have introduced a physical mechanism that explains the electrophysiological death of neurons due to the influence of A$\beta$ oligomers in the context of AD. We note that some of the key features of our analysis agree with the findings from other related and direct observational studies.  Thus it is found by electrophysiological measurements in cultured hippocampal neurons that elevated A$\beta$ production disrupts both presynaptic and postsynaptic functions \cite{Ting07}.  Our results also indicate that synaptic depression occurs in all the neurons.  A study of  hippocampal neurons reported that the synaptic depression requires neuronal activity \cite{Kam03}. In our model also, with no neuronal activity, A$\beta$ will not become electrically active and therefore cannot cause depression. We observe that the synaptic silencing is associated with considerable disruption of synchronized activity.  This is in agreement with the reported disruption of cortical synaptic integration in the presence of A$\beta$ in transgenic mice\cite{Ste04}. 

Some of the mechanisms for AD reported recently are either biochemical due to generation of free radicals \cite{Var00} or chemical due to binding of A$\beta$ in cell membrane \cite{Bla02,Sok06}.  The mechanism introduced in this paper is novel in that it is the first electrical model to the best of our knowledge. Moreover, our model for the first time proposes the presence of induced activity in A$\beta$ oligomers when synaptic disruption occurs. We note that indirect evidence for induced electrical activity can be drawn from the reported studies on birefringence in certain amyloid proteins such as $\beta$-Lactoglobulin under applied electrical pulses \cite{Rog06}.  This indicates the possibility of such an activity in A$\beta$ under the influence of neuronal signals. A direct test for such induced activity can be obtained by electrophysiological measurements in cultured media or brain slices.  Such experiments can also lead to possible revival strategies by external interventions of the type proposed in our model.
 
None of the studies reported so far explain precisely the reason why A$\beta$ monomer is not effective in inducing synaptic suppression but dimers and other n-oligomers are.  We conjecture that a possible explanation through our model would be that the monomer does not develop electrical activity of the required magnitude to have a measurable effect, while higher oligomers develop electrical activity of the same order of magnitude as a normal synaptic interaction.  A direct measurement of the induced effect on amyloid monomer and dimer in a field equivalent to that across the synapse can give direct support for this conjecture. Thus the theory proposed here opens up interesting observational research in at least two directions. 
We have obtained qualitatively similar results with other models such as Hodgkin-Huxley and FitzHugh- Nagumo models. These results are presented in the supporting information. Finally, we propose that the mechanism introduced here is quite general and can be applied in the context of other neurodegenerative diseases such as Parkinson's disease, Huntington's disease, Creutzfeld- Jakob disease, etc where specific protein deposits have been identified as the neurotoxic agents \cite{Cau03}.

\end{document}